\documentclass[prb,twocolumn,superscriptaddress]{revtex4-2}

\usepackage{amsmath,amssymb,mathrsfs}
\usepackage{graphicx}
\usepackage{colortbl}
\usepackage{braket}
\usepackage{bm}
\usepackage{amsfonts}
\usepackage{ulem}
\usepackage[colorlinks,plainpages=false,linkcolor=blue,urlcolor=blue,citecolor=blue,pdfpagemode=UseNone,pdfstartview=FitBH]{hyperref}

\begin{document}
\title{Giant bubbles of Fisher zeros in the quantum XY chain}

\author{Songtai Lv}
\affiliation{Key Laboratory of Polar Materials and Devices (MOE), School of Physics, East China Normal University, Shanghai 200241, China}

\author{Yang Liu}
\altaffiliation{yliu1130@ucr.edu}
\affiliation{Department of Physics and Astronomy, University of California, Riverside, California 92521, USA}

\author{Erhai Zhao}
\altaffiliation{ezhao2@gmu.edu}
\affiliation{Department of Physics and Astronomy, George Mason University, Fairfax, Virginia 22030, USA}

\author{Haiyuan Zou}
\altaffiliation{hyzou@phy.ecnu.edu.cn}
\affiliation{Key Laboratory of Polar Materials and Devices (MOE), School of Physics, East China Normal University, Shanghai 200241, China}
\affiliation{State Key Laboratory of Low-Dimensional Quantum Physics, Department of Physics, Tsinghua University, Beijing 100084, China}

\author{Tao Xiang}
\altaffiliation{txiang@iphy.ac.cn}
\affiliation{Beijing National Laboratory for Condensed Matter Physics and Institute of Physics,
Chinese Academy of Sciences, Beijing 100190, China}
\affiliation{School of Physical Sciences, University of Chinese Academy of Sciences, Beijing 100049, China}

\begin{abstract}
We demonstrate an alternative approach based on a complex-valued inverse temperature and partition function 
to probe quantum phases of matter with nontrivial spectra and dynamics. It leverages 
thermofield dynamics (TFD) to quantitatively characterize quantum and thermal fluctuations, and 
exploit the correspondence between low-energy excitations and Fisher zeros. 
Using the quantum XY chain in an external field as a testbed, we 
show that the oscillatory gap behavior manifests as oscillations in the long-time dynamics of the TFD spectral form factor.
We also identify giant bubbles, i.e. large-scale closed lines, of Fisher-zeros 
near the gapless XX limit. They provide a characteristic energy scale that seems to contradict the predictions of the low energy
theory of a featureless Luttinger liquid. We identify this energy scale and relate the motion of these giant bubbles with varying external field to
the transfer of spectral weight from high to low energies. The deep connection between Fisher zeros, dynamics, and excitations opens up promising avenues for understanding the unconventional gap behaviors in strongly correlated many-body systems.
\end{abstract}

\maketitle

\textit{Introduction.} {The partition function} is one of the central objects in statistical mechanics. Compared to spectral or correlation functions
that directly relate to experimental observations, it receives less attention in conventional theoretical approaches to strongly correlated quantum systems.
Some of these systems refuse to fall into the paradigms of order parameters and quasiparticles
and therefore call for new perspectives or tools. One striking example is the so-called pseudogap in high-temperature superconductors~\cite{Zaanen2015,Timusk1999,TALLON200153,Lee2006}, which is related to the lack of consensus on the phase diagram of a two-dimensional (2D) Hubbard model and related strongly correlated lattice models~\cite{Wen1996,Varma2006,Zhang2006,Qin2020,Xu2024}. 
Recently, fueled by the growing interest in open and monitored quantum systems~\cite{fisher2023random,Li2018prb,Skinner2019prx,Chan2019prb,Basu2022prr,Jian2021,Matsumoto2022}, an ``unconventional" probe of quantum many body systems has been proposed. It analytically continues the partition function $Z$ to be a function of the complex inverse temperature $\beta =\beta_r+i\beta_i\in \mathbb{C}$. Roughly speaking, $Z$ accounts for both thermal and quantum fluctuations, with $1/\beta_r$ playing the role of temperature and $\beta_i$ as time.

The idea of complex $\beta$ goes back to Fisher who, following the pioneering work of Lee and Yang~\cite{LeeYang1952I,LeeYang1952II}, showed that the location of zeros of $Z$ on the complex $\beta$ plane, 
as the solution of $Z(\beta=\beta_r+i\beta_i)=0$ and now referred to as Fisher zeros, can be used to identify thermodynamic phase transitions~\cite{Fisher1965statistical}.
Later, in the study of quench dynamics from a given pure state, the zeros of the so-called boundary partition function (analogous to $Z$) were used to describe dynamical quantum phase transitions~\cite{DynamicalPT2013PRL}. 
We have shown that the complex $Z$ and its Fisher zeros can be applied to quantum many-body systems to 
characterize quantum criticality~\cite{liu2024PRR,Liu2024CPL}, low-energy excitations~\cite{liu2023CPL}, breakdowns of the eigenstate thermalization hypothesis~\cite{Meng2025a}, and even topological features~\cite{Meng2025b}. By now, there is a surge of interest in complexified quantum many-body systems~\cite{Flindt2021,liu2023CPL,Ueda2023prl,Li2024pre,liu2024PRR,Liu2024CPL,Chen2024prb,Wada2025LY, Yang2025prb,Meng2025a,Meng2025b,Shastry2025pre,JWang2025prb,Li2025,Sun2026,Zheng2026prbl}. In this context, the best understood example of Fisher zeros 
is the the one-dimensional transverse field Ising model (TFIM). For example, an open line of Fisher zeros extending to $\beta_r\rightarrow\infty$
in the complex $\beta$-plane indicates low energy excitations, such as domain-walls, at energy scale set by $1/\beta_i$. {A natural question then is
what happens to the configuration of Fisher zeros if a system does not have well-defined quasiparticles but an enhanced, broad, incoherent spectral weight (the suppressed spectral weight then qualifies as a pseudogap)? Clearly, they should not form open lines. To address these questions, we need to go beyond TFIM to study models that feature interesting, unconventional phases that do not resemble a Fermi liquid. }

\begin{figure}[t]
\includegraphics[width=0.4\textwidth]{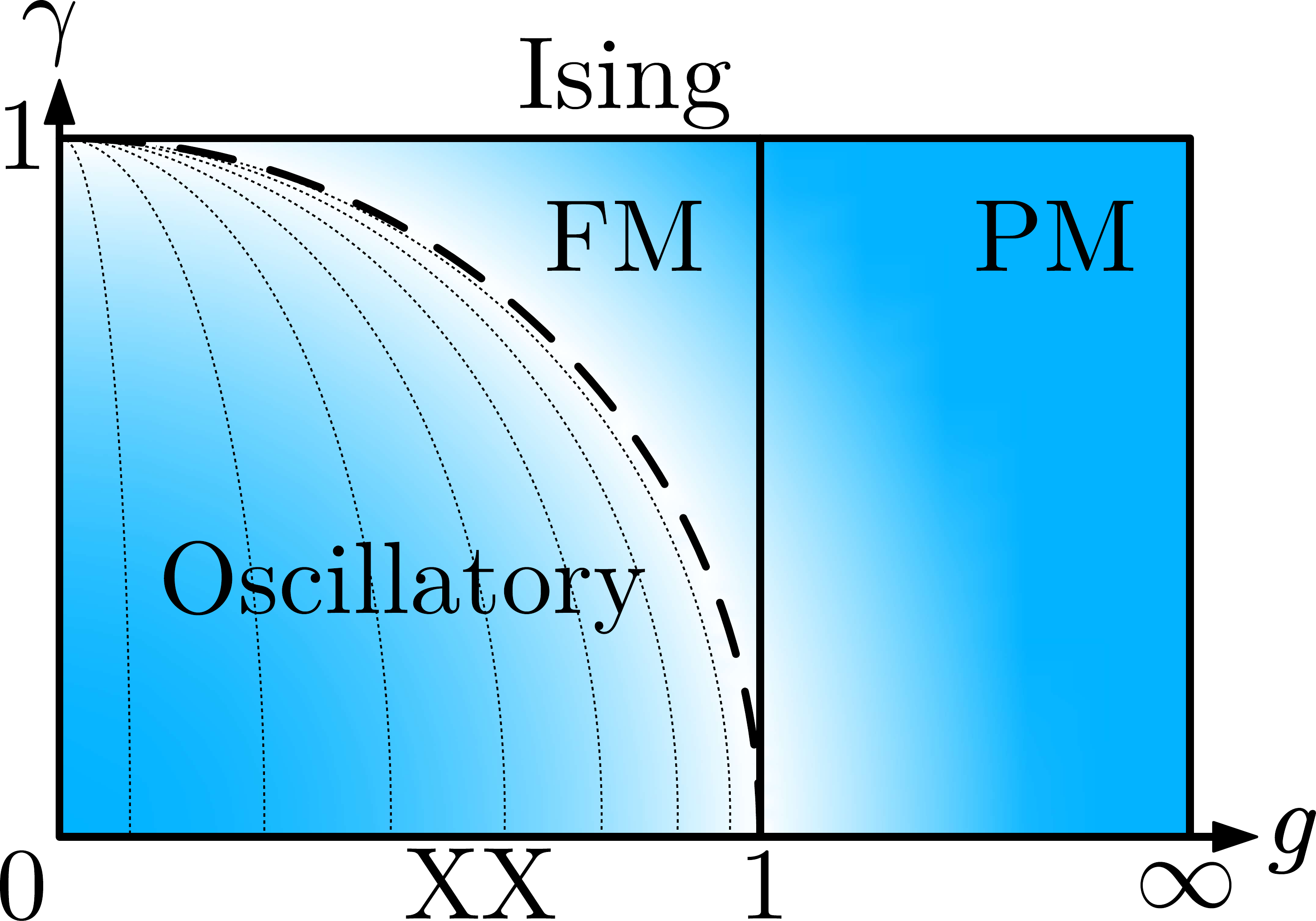}
\caption{
Phase diagram of the quantum XY model under an external field $g$. $g=1$ is the critical line separating the FM phase from the {PM} phase. The thick dashed line separates the oscillatory gap region from the FM region. The thin dashed line indicates where the gap vanishes for $L = 16$. $\gamma=1$ corresponds to the Ising limit, while $\gamma=0$ represents the gapless XX limit.
}
\label{fig:fig1}
\end{figure}

These considerations motivated us to investigate the complex partition function and Fisher zeros for the 1D XY model, Eq. \eqref{Hamiltonian}. {This quantum spin system has several appealing features.} First, it contains TFIM as a special limit which serves as a sanity check to ensure the result is correct. Second, $Z$ can be computed exactly. Third, the model has a rich phase diagram including a ferromagnetic {(FM)} phase, a rather peculiar oscillatory phase, a gapless Luttinger liquid, and {paramagnetic (PM)}  phase. 
Finally, it exhibits rich spectral structures at both low and high energies. We shall show a few surprising behaviors (see Fig. \ref{fig:fig4}) of Fisher zeros as the model parameters are tuned, e.g., approaching various phase transitions.

\textit{Unpacking the complex partition function $Z$.} {Before introducing the XY model, we first recap the complex partition function can be viewed from the perspective of thermofield dynamics~\cite{takahashi1996thermo}. This will set the stage for subsequent technical discussions.}
If we restrict $\beta$ to be purely imaginary, then the modulus square of $Z$ is known as the spectral form factor $S(t)=|Z(\beta=it)|^2$, a popular 
device in quantum chaos and random matrix theory to characterize energy level statistics~\cite{cotler2017black,delCampo2017,YLiao2020}. Thus spectral information is baked into $Z$ once we introduce the imaginary part of $\beta$.
For general $\beta$, we view $Z$ as the return amplitude of a pure state once we purify the thermal ensemble. This can be done by introducing 
the thermofield double (TFD) state $|\Psi(\beta_r,0)\rangle = \sum_n e^{-\beta_r E_n/2}|n\rangle_L\otimes|n\rangle_R$ where the degrees of freedom have been doubled (the original $L$ and its copy $R$)~\cite{takahashi1996thermo}. Then $Z(\beta_r,t)\sim \langle\Psi(\beta_r,0)|\Psi(\beta_r,t)\rangle$, where the inverse temperature $\beta_r$ serves as a parameter while  $t=\beta_i$ is the real time. 
At large $\beta_r$ {and in the thermodynamic limit, the dominant contribution to the oscillatory part of the normalized $S=|Z(\beta=\beta_r+i\beta_i)/Z(\beta=\beta_r)|^2$ comes from $2e^{-\beta_r\Delta_\infty}\cos(\Delta_\infty\beta_i)$, where $\Delta_\infty$ is the lowest excitation gap in the thermodynamic limit. For the XY model with a finite system size $L$, in contrast, if $L$ is kept finite and $\beta_r$ is increased, the long-time dynamics of $S$ will be dominated by $2e^{-\beta_r\Delta_L}\cos(\Delta_L\beta_i)$. Here $\Delta_L$ is a finite-size fine-structure gap satisfying $\Delta_L\ll \Delta_\infty$ (If $L$ is taken to infinity first, $\Delta_L$ becomes negligible)~\cite{supp}.}
Therefore, by choosing appropriate values of $L$, $\beta_r$, and $\beta_i$, one can access both the thermodynamic-limit and finite-size gap structures. A careful treatment of these limits is crucial to discuss the oscillatory phase of the XY model.

With the advancement in tensor network algorithms, the complex $Z$ can be computed reliably for lattice spin models~\cite{XieHOTRG,Zou2014PRD} to offer 
a unified framework to capture the competition between thermal and quantum fluctuations. In this Letter, we shall use the exact solutions of complex $Z$,
in particular the asymptotic behaviors of $S$ and the evolution of Fisher zeros, to diagnose the quantum XY model. {Instead of directly solving for the roots of the equation, the exact numerical contour solution (obtained by adjusting the precision to locate the sign change of the function) is used to determine the Fisher zeros.} {Our main objective is to find out whether these theoretical devices yield additional insights about more ``exotic" (the incommensurate phase with oscillating finite-size gaps) or ``featureless" Luttinger liquids (which turn out to have rather interesting interplays of low and high energy excitations) beyond TFIM.}

\begin{figure}[t]
\includegraphics[width=0.5\textwidth]{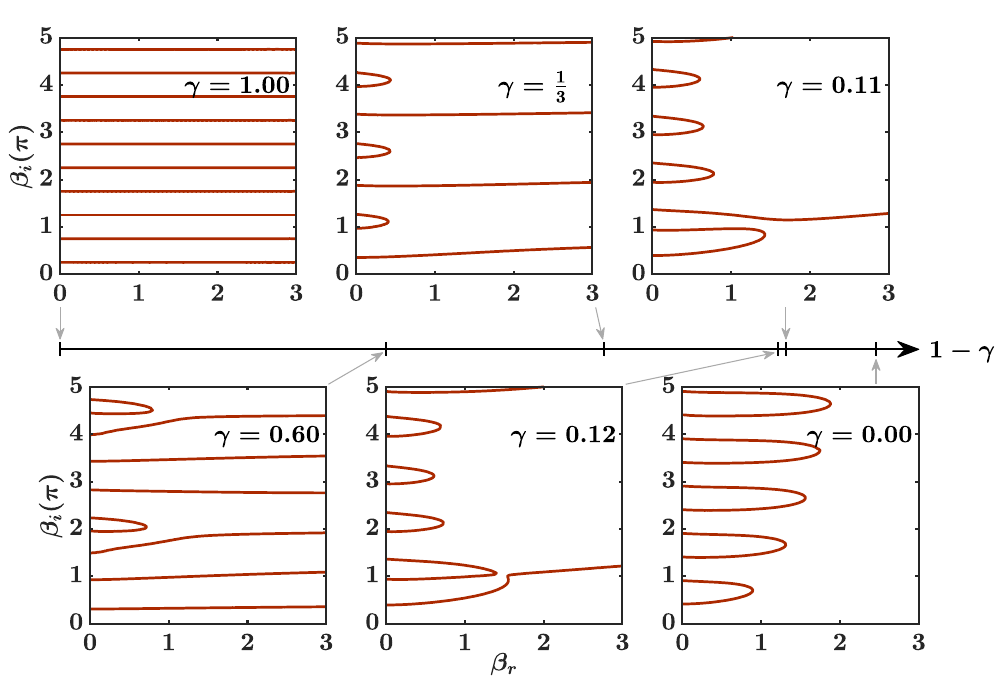}
\caption{
Fisher zeros of the XY model without external field ($g = 0$) in the thermodynamic limit for different values of $\gamma$. As $\gamma$ decreases from the Ising limit ($\gamma = 1$) toward the XX limit ($\gamma = 0$), an upward-moving open Fisher zero line crosses over with a downward-moving closed Fisher zero line. The cases $\gamma = 0.12$ and 0.11 provide the crossover example closest to the $\beta_r$-axis.
}
\label{fig:fig2}
\end{figure}

\begin{figure*}[htbp]
\includegraphics[width=\textwidth]{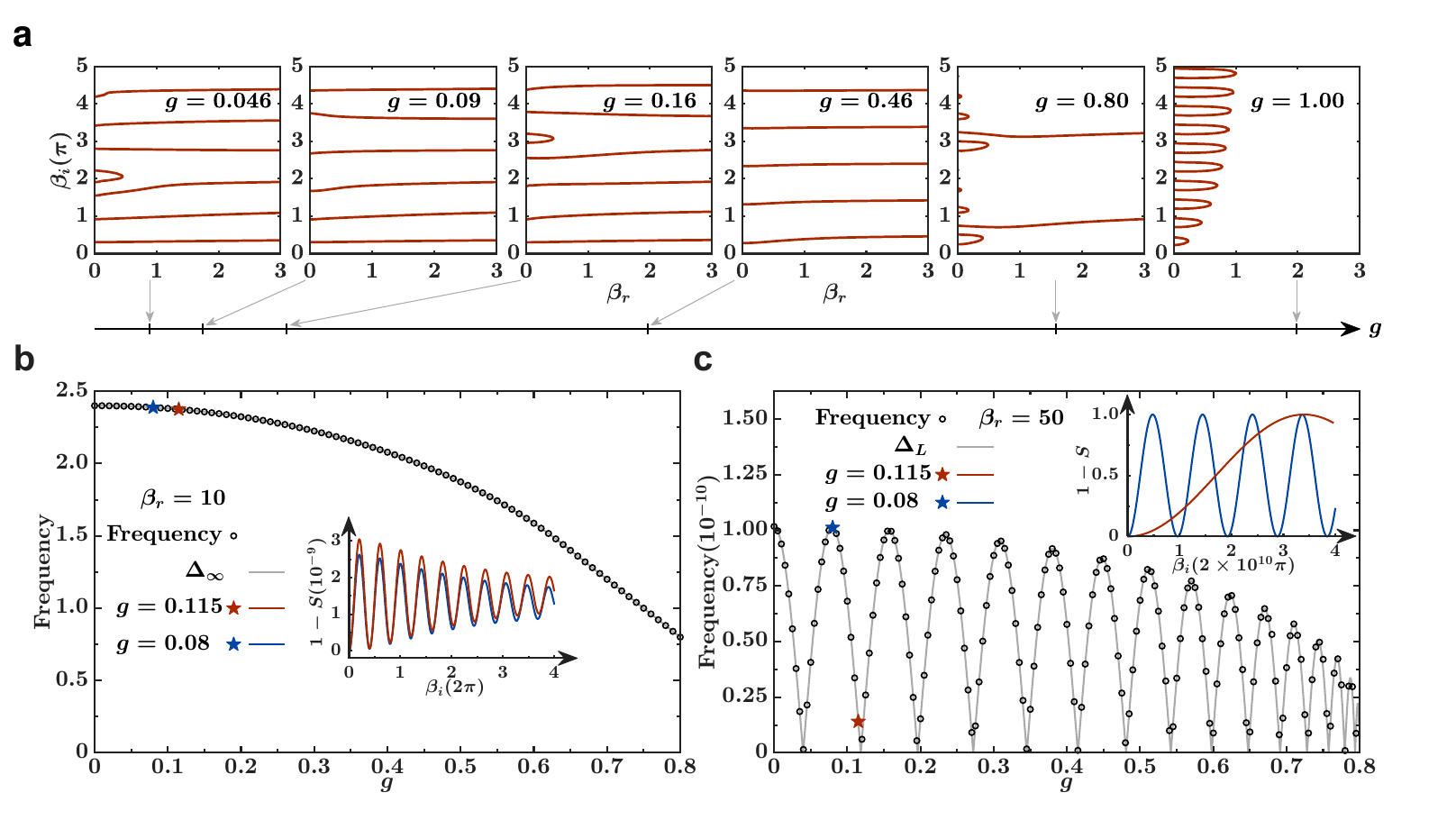}
\caption{
    Oscillatory behavior of Fisher zeros and $S$ for $\gamma = 0.6$. (a) Evolution of Fisher zero configurations in the thermodynamic limit as $g$ increases. Before reaching $g=1$, the closed Fisher zero loops exhibit non-monotonic behavior, including disappearing and reappearing at $g\lesssim 0.5$. 
    (b) For $L=32$ and $\beta_r=10$, the oscillation frequency of $S$ (black open dots) agrees well with $\Delta_\infty$ (gray line). The inset shows the oscillation of $1-S$ with $\beta_i$ for two nearby values $g=0.08$ (blue pentagram) and $g=0.115$(red pentagram). (c) For $L=32$ and $\beta_r=50$, the oscillation frequency of $S$ (black open dots) matches $\Delta_L$ (gray line). This inset shows the long-time oscillatory behavior of $1-S$ for the two $g$ values from (b). 
}
\label{fig:fig3}
\end{figure*}

\textit{{Exact Fisher zeros for} the 1D quantum XY model.} 
The Hamiltonian of the 1D $XY$ model reads 
\begin{equation}
    H = - \! \sum_{j=1}^{L} \! \left[ 
        \frac{1+\gamma}{2}\sigma_{j}^{x} \sigma_{j+1}^{x} 
        \!+\! \frac{1-\gamma}{2} \sigma_{j}^{y} \sigma_{j+1}^{y} 
        \!+\! g \sigma_{j}^{z}
     \right]\!.
    \label{Hamiltonian}
\end{equation}
Here $\sigma^\mu_j~(\mu = x, y, z)$ are Pauli matrices at site $j$, $g$ is the transverse field, $\gamma$ is the anisotropy parameter, and period boundary condition is assumed. As a paradigmatic example of quantum magnetism, this model has a rich phase diagram~\cite{Barouch1971,Hoeger1985,Okuyama2015} summarized in Fig.~\ref{fig:fig1}. {The $g=1$ line separates the {PM phase} ($g>1$) from the ordered FM phase ($g<1$). But the region bounded by the line $\gamma^2 + g^2 = 1$ is peculiar: its spin correlation function oscillates on top of power law decay, and for finite system sizes, the gap also exhibit oscillations.} As illustrated in Fig.~\ref{fig:fig1} for $L=16$, {the gap vanishes along a set of} lines in the $g-\gamma$ plane, which become increasingly dense with larger system sizes~\cite{Okuyama2015}. {This region is usually referred to as the oscillatory or incommensurate phase. The model reduces to TFIM along the line $\gamma=1$, and for $\gamma=0$, it reduces to the XX model} which serves as a representative example of a gapless Luttinger liquid
in the thermodynamic limit.

Our analysis begins by analytically continuing the exact partition function $Z$ of the XY model to complex values of the inverse temperature $\beta$. Note that the exact form of $Z$ is subtle and has not been clarified until recently~\cite{Biaoczyk2021,Franchini2017}. For $g<1$, by extending the technique we used earlier to obtain the Fisher zeros of TFIM~\cite{liu2023CPL}, we can derive the exact Fisher zeros of the XY model off the $\beta_i$-axis in the thermodynamic limit. They are given by the transcendental equation{~\cite{supp}}
\begin{equation}
    \int_{0}^{\pi} \log|\tanh(\beta \epsilon_q)|^2 {\rm d}{q} = 0,
    \label{continuousZero}
\end{equation}
where $\epsilon_q = \sqrt{[g - \cos(q)]^2 + [\gamma \sin(q)]^2}$. {Eq. \eqref{continuousZero} is exact and one of the main results of this work.}

{To appreciate the form and location of Fisher zeros, let us first examine the case of zero transverse field, $g=0$, as shown in Fig.~\ref{fig:fig2}. 
We observe that similar to the case of TFIM~\cite{liu2023CPL}, the zeros congregate either to open lines or closed loops, or ``bubbles" (only half the loops are shown as the diagram is mirror symmetric with respect to the $\beta_I$ axis). The open lines extend to large $\beta_r$ (low temperatures), and as we have shown earlier~\cite{liu2023CPL}, its vertical location, i.e. the inverse of the imaginary part of the Fisher zeros ($1/\beta_i$), provides a useful diagnostic measure for the lowest energy gap of the system.} As $\gamma$ is decreased (Fig.~\ref{fig:fig2}), the open Fisher zero lines shift upward in the complex $\beta$ plane 
accompanied by the downward movement of closed zeros. {In other words, the lines rise while the bubbles sink.} This is in accordance with the closing of the gap as $\gamma$ decreases. In the limit of $\gamma = 0$, the system becomes gapless and the open zero lines vanish completely. {This example provides another illustration that the structure change of Fisher zeros can effectively pinpoint} quantum criticality.

\begin{figure*}[htbp]
\includegraphics[width=1.0\textwidth]{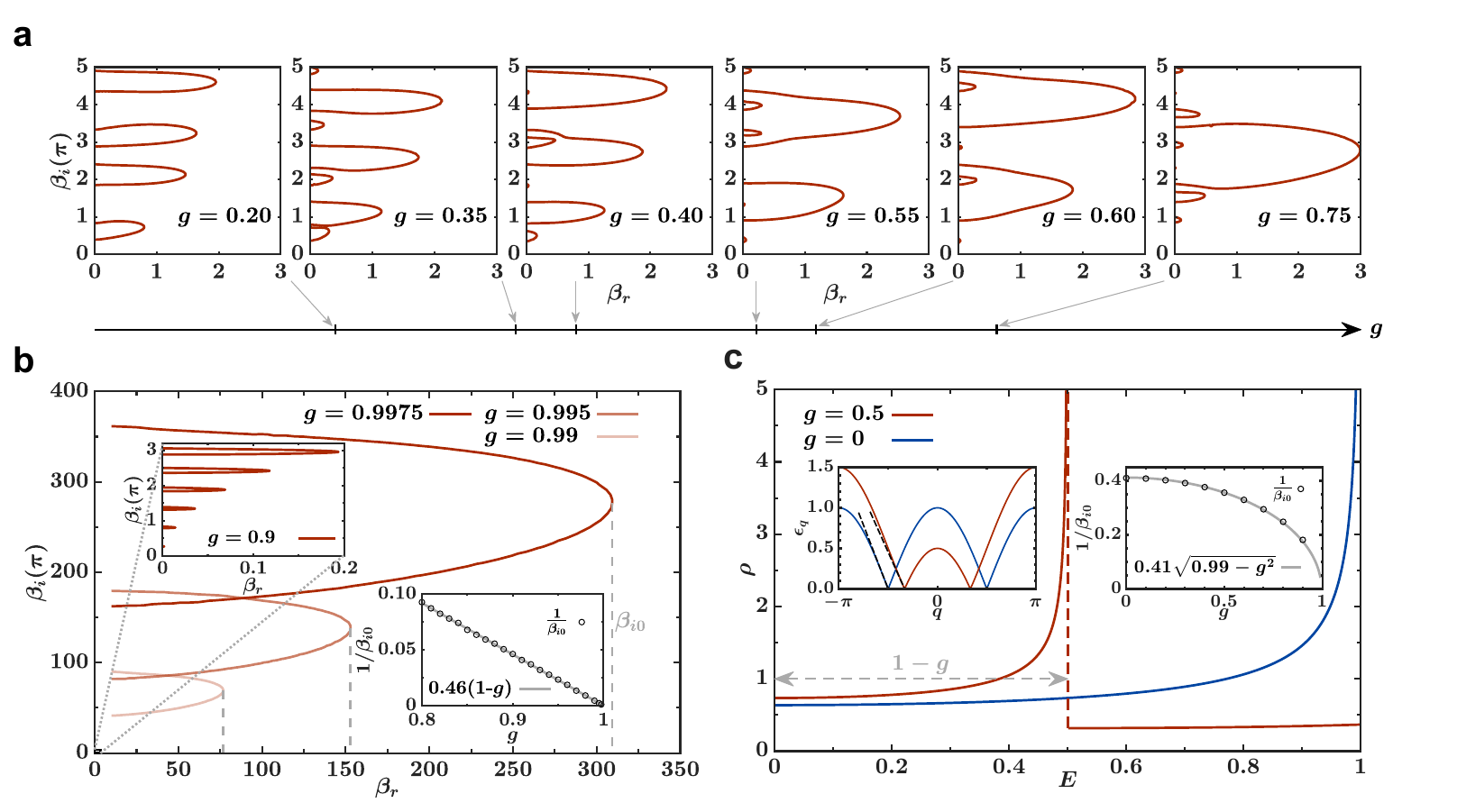}
\caption{
   Large-scale closed Fisher zero patterns at $\gamma=0$. (a) For different $g$s, the Fisher zeros form only closed loops. As $g$ increases, the closed zero loops exhibit a tendency to expand. (b) Near the QCP, the expansion of the large-scale closed Fisher zero loop becomes more pronounced compared to the smaller zero loops near the $\beta_i$-axis (upper-left inset). The lower-right inset shows that the inverse of the imaginary part ($1/\beta_{i0}$) of the rightmost Fisher zero on the large loop scales linearly with $1-g$. (c) The DOS diverges near $1-g$, as illustrated for $g=0$ (blue) and $g=0.7$ (red), exhibiting van Hove singularities. The left inset shows the corresponding dispersion relations $\epsilon(q)$, where the dashed slope illustrates the Luttinger-liquid velocity, which is proportional to $\sqrt{1-g^2}$. The right inset displays the dependence of $1/\beta_{i0}$ of the open zero lines on $g$ ($1/\beta_{i0}\sim\sqrt{0.99-g^2}$) at $\beta_r=100$ when a small anisotropy $\gamma=0.1$ is introduced.}
\label{fig:fig4}
\end{figure*}

\textit{{Short- and long-time dynamics in the oscillatory phase.}} To see whether the complex $Z$ can shed some light on the enigmatic oscillatory phase within $\gamma^2+g^2<1$, we show in Fig.~\ref{fig:fig3}(a) the evolution of Fisher zeros along the cut $\gamma=0.6$. 
Here the overall trend is that increasing $g$ leads to an upward shift of the open zero lines and a downward movement of the closed loops (bubbles). However, in contrast to Fig.~\ref{fig:fig2}, the motion of the bubbles at small $\beta_r$ values is non-monotonic: they can appear or disappear suddenly. This anomalous behavior, which is absent in TFIM, can be taken as another marker of the oscillatory phase. In other words, the oscillatory phase leaves its imprint on the Fisher zero configurations in the thermodynamic limit.

{To establish a direct and quantitative connection to the oscillatory gap behavior (dotted lines in Fig.~\ref{fig:fig1}), we
further compute the normalized $S$ for finite-size systems across the entire parameter range of $\gamma>0$. As discussed earlier, both 
the finite-size gap ($\Delta_L$) and the gap in the thermodynamic limit ($\Delta_\infty$) can be extracted by taking different limits of $\beta$ and $L$.} 
In Fig.~\ref{fig:fig3}(b), we choose $\beta_r=10$ and $L=32$ for $\gamma=0.6$. Since $\beta_i$ plays the role of time, the short-time oscillation frequency of $S$ gives $\Delta_\infty$ which is in good agreement with the analytical result 
$\Delta_{\infty}=4\gamma\sqrt{(1-\gamma^2-g^2)/(1-\gamma^2)}$ for $g<1-\gamma^2$, and by $|1-g|$ for $g>1-\gamma^2$
determined from the condition $d\epsilon_q/dq=0$~\cite{Wojtkiewicz2016}. This agreement becomes even better as both $L$ and $\beta_r$ increase. 
In this regime, adjacent values of $g$ have only minor differences in their oscillation periods.
The finite-size gap $\Delta_L$ is extracted by fixing $L$ using significantly larger $\beta_r$. An example is shown in Fig.~\ref{fig:fig3}(c),
where huge values of $\beta_i$ are used to show the long-time oscillation of $S$. 
{If $\beta_r$ is fixed and $L$ is increased, the oscillation frequency decreases and approaches zero in the thermodynamic limit.}
The extracted $\Delta_L$ oscillates with $g$, in excellent agreement with the exact solutions for $\Delta_L$ reported in the literature~\cite{Okuyama2015}. 
In this case, even small differences in $g$ lead to noticeable changes in the oscillation frequencies of $S$ (see inset). 
{Thus, the thermofield dynamics encoded in $S$ provides unambiguous evidence to identify the oscillatory phase in the XY model.
This is the second main result.}

\textit{{Giant Fisher bubbles.}} Now let us focus on the XX limit at $\gamma=0$, 
where the system remains gapless for $g\leq 1$. What do the Fisher zeros look like in this phase? 
The complex $\beta$-plane must contain no open zero lines; otherwise, as $\beta_r\rightarrow\infty$, the zeros would indicate a finite gap, contradicting the gapless nature. The exact solutions of Fisher zeros in the thermodynamic limit confirm this argument. 
{Moreover, it shows that all zeros are in the form of closed bubbles,} see
Fig.~\ref{fig:fig4}(a). {An interesting trend emerges as $g$ is increased:} in addition to a population of small bubbles near the $\beta_i$-axis, 
some bubbles grow in size by extending toward larger and larger $\beta_r$. As the bicritical point ($g=1$, $\gamma=0$) is approached, giant bubbles of Fisher zeros emerge to dominate the scene [Fig.~\ref{fig:fig4}(b)].
Its size $\beta_{i0}$, as measured by the imaginary part of its rightmost endpoint, diverges as $g\rightarrow 1$. 
This implies that at very low temperatures, the system {exhibit a ``pseudogap" scale $\sim 1/\beta_{i0}$}. 
The inset of Fig.~\ref{fig:fig4}(b) shows that $1/\beta_{i0}$ scales linearly with $1-g$. {The inverse of the fitting coefficient, 0.46, corresponds to the imaginary-axis ``fixed-point" coordinate of the rescaled variable $\tilde{\beta}=(1-g)\beta$ as $g\rightarrow 1$~\cite{supp}}. What sets this energy scale
with enhanced spectral weight?

The low-energy physics of the XX model at $\gamma=0$ can be effectively described by a Luttinger liquid
with a linear dispersion $\epsilon(k)=v_F(k-k_F)$ around the Fermi points. {This theory does not possess 
any characteristic energy scale other than the energy cutoff. To understand the pseudogap scale, we must go beyond the Luttinger description by considering
the full energy-momentum dispersion $\epsilon_q$ of quasiparticles which are collective excitations of interacting spins.} 
As shown in Fig.~\ref{fig:fig4}(c), the band-edge extrema as a part of the ``high-energy" physics give rise to van Hove singularities, i.e. divergence in the density of states (DOS) $\rho(E)=\sum_{q:\epsilon_q=E}|d\epsilon_q/dq|^{-1}/2\pi$, at energy $\epsilon_0=(1-g)$. 
This scaling behavior agrees with $1/\beta_{i0}\propto (1-g)$ observed from Fig.~\ref{fig:fig4}(b). Therefore, we can trace the origin of the pseudogap to 
the van Hove singularity. When $g\rightarrow 1$, the Fermi points move toward the band edge at small $q$. The result is a crossover where the separation between low- and high-energy features becomes ill-defined. The van Hove singularity merges with Fermi points with vanishing Fermi velocity $v_F\sim\sqrt{1-g^2}$. On the complex plane, this process manifests itself as giant bubbles of Fisher zeros of ever increasing size.

{To further check the validity of this interpretation, we investigate what happens at small $\gamma$ when the system is 
driven away from the XX limit.} Finite $\gamma$ opens a gap, $\Delta_{\inf}\sim\gamma v_F$.
Thus, we predict that the open Fisher line associated with gap bends back and crosses over with the giant bubble. Moreover,  The size $1/\beta_{i0}$ of this open Fisher line should scale with $v_F\propto \sqrt{1-g^2}$ at large $\beta_r$. 
This is indeed what we observe from the numerical fit in the right inset of Fig.~\ref{fig:fig4}(c), where $\gamma=0.1$ and $\beta_r=100$. 
{Note that these giant bubbles are absent in TFIM. The theoretical discovery and understanding of these giant bubbles near the XX limit, including their
correlation and scaling with characteristic energy scales in the spin system, are the third main results of this work.}

\textit{Summary and outlook.} 
{Our detailed analysis confirms} that complex partition function offers an effective tool for analyzing the thermal and dynamical properties
of the 1D $XY$ model. By varying $\beta_r$ and $\beta_i$ and investigating the Fisher zeros and thermofield dynamics (through $S$), we 
accurately capture the oscillatory finite-size gap behavior. 
In the thermodynamic limit, we uncover a distinctive class of giant bubbles
(large-scale closed lines of Fisher zeros) near the critical line. 
{We further demonstrate that the evolution of these bubbles with varying external field 
offers an intuitive yet sensitive probe for the transfer of spectral weight within the Luttinger liquid phase.}
Since the pseudogap phenomenon~\cite{Tsvelik2000,CHEN20051,Mueller2017,Chen2024rmp} is characterized by a 
redistribution of the spectral weight, we conjecture that 
Fisher zeros and thermofield dynamics may offer a different perspective for exploring the 
intricate many-body physics related to unconventional gap behaviors. 
{The scaling of energy levels, inferred from the inverse of the imaginary part of the zeros, yields the critical exponent $\nu=1$ in the Ising limit. A worthwhile direction for future work is to further exploit finite-size scaling and the distribution and density of zeros~\mbox{\cite{Itzykson1983,Janke2004,Sumaryada2007}} in various zero configurations to uncover more critical behavior.
As Lee-Yang zeros have begun to be explored experimentally~\cite{Peng2015}, and with the advent of methods such as nonunitary evolution in quantum circuits ~\mbox{\cite{Francis2021SA,Jafferis2022,fisher2023random,Granet2023prl}}, Rydberg atom systems~\mbox{\cite{Lukin2017Nature,You2024NP}}, and trapped ions~\cite{Zhu2020}, the experimental observation of Fisher zeros is expected to become feasible.}
Our approach outlined here can be extended to 2D strongly interacting models for which 
the complex-valued partition functions cannot be solved exactly but can be evaluated to high precision by tensor network algorithms~\cite{XieHOTRG,Zou2014PRD}.

{~\it Acknowledgments.} We thank Jiansong Pan, Fan Yang, and Xuefeng Zhang for helpful discussions. H.Z. is supported by the National Natural Science Foundation of China (Grant No. 12274126) and Open Research Fund Program of the State Key Laboratory of Low-Dimensional Quantum Physics. E.Z. acknowledges the support from NSF Grant PHY-206419. 

{~\it Data availability.} The data that support the findings of this article are openly available~\cite{lv2026zenodo}.

%

\newpage

\widetext
\begin{center}
\textbf{\large Supplementary material for \\``Giant bubbles of Fisher zeros in the quantum XY chain"}
\end{center}

\setcounter{equation}{0}
\setcounter{figure}{0}
\setcounter{table}{0}

\makeatletter
\renewcommand{\thefigure}{S\arabic{figure}}
\renewcommand{\thetable}{S\arabic{table}}
\renewcommand{\theequation}{S\arabic{equation}}
\renewcommand{\bibnumfmt}[1]{[S#1]}
\renewcommand{\citenumfont}[1]{S#1}
\renewcommand{\bibsection}{} 
\makeatother

\section{Detail of the exact Fisher zeros in the thermodynamic limit}

The exact form of the partition function $Z$ of the XY model is described as ~\cite{SBiaoczyk2021,SFranchini2017,Sliu2024PRR}:
\begin{align}
	Z = \frac{1}{2}
	\left[
	\prod_{k=1}^L 2\cosh\left(\beta\epsilon_{2k}\right)+\mathrm{sign} (g_c-g)\prod_{k=1}^L 2\sinh(\beta\epsilon_{2k}) + \prod_{k=1}^L 2\cosh(\beta\epsilon_{2k-1})+\prod_{k=1}^L 2\sinh(\beta\epsilon_{2k-1})
	\right]
	\label{eq.XY.exactZ}
\end{align}
where $L$ is the system size, $\epsilon_k = \sqrt{[g-\cos(k\pi/L)]^2 + [\gamma \sin(k\pi/L)]^2}$, $g_c=1$ is the critical field, and $\beta$ is the inverse temperature.

For $g < g_c$, the exact Fisher zeros of the XY model in the thermodynamic limit ($L \to \infty$) can be derived for regions away from the imaginary axis of the complex temperature ($\text{Re}~\beta \neq 0$) by extending techniques previously developed for the transverse-field Ising model~\cite{Sliu2023CPL}. Starting from the finite-lattice exact solution Eq. ~\ref{eq.XY.exactZ}, the quasiparticle energy levels for adjacent momentum indices satisfy $\epsilon_{2k} \approx \epsilon_{2k-1}$ when $L$ is sufficiently large. Consequently, the even and odd parity sectors in Eq. \ref{eq.XY.exactZ} coincide and their difference does not contribute in the thermodynamic limit. The partition function simplifies to a combination of two leading products. The condition $Z=0$ then becomes equivalent to:
\begin{equation}
    \prod_{k=1}^L 2\cosh(\beta \epsilon_{2k}) + \prod_{k=1}^L 2\sinh(\beta \epsilon_{2k}) = 0.
\end{equation}
Dividing both sides by the first term yields:
\begin{equation}
	\label{eq.1+tanh=0}
    \prod_{k=1}^L \tanh(\beta \epsilon_{2k}) = -1.
\end{equation}

Taking the squared modulus, then the logarithm, and finally dividing by $L$ yields
\begin{align}
  \label{eq.finitk}
  \frac{1}{L} \sum_{k=1}^{L} \log |\tanh(\beta\epsilon_{2k})|^2 = 0.
\end{align}

In the thermodynamic limit $L \to \infty$, 
by setting $q=2k\pi/L$, Eq.~\ref{eq.finitk} can be regularized to a continuous form. Then, exploiting the symmetry of the integrand over the intervals $[0,\pi]$ and $[\pi,2\pi]$ ($\epsilon_q = \epsilon_{2\pi-q}$), we obtain Eq.~(2) in the main text:
\begin{align}
  \int_{0}^{\pi}  \log |\tanh(\beta\epsilon_q)|^2 \text{d} q =0,
\end{align}
with $\epsilon_q = \sqrt{(g-\cos q)^2 + (\gamma \sin q)^2}$. 

\section{Coexistence of $\Delta_\infty$ and $\Delta_L$ in the spectral form factor}

This section elucidates that the real-time oscillations in the spectral form factor $S$ at frequency $\Delta_\infty$ arise from excitations within each parity sector of the partition function $Z(\beta=\beta_r+i\beta_i)$, whereas those at frequency $\Delta_L$ correspond to the vacuum energy difference between distinct parity sectors. The total partition function is partitioned into two parity sectors as
\begin{equation}
    Z = Z_1 + Z_0,
\end{equation}
where the index $p \in \{0, 1\}$ distinguishes the even and odd sectors, respectively. The energy spectra within each sector are denoted as $\epsilon_k^{(p)}$, with $\epsilon_k^{(1)} = \epsilon_{2k}$ and $\epsilon_k^{(0)} = \epsilon_{2k-1}$. In the ferromagnetic phase $g < g_c$, the contribution of each sector to the partition function is defined as
\begin{align}
    Z_p &= \frac{1}{2} \left[ \prod_{k=1}^L 2\cosh (\beta\epsilon_k^{(p)}) + \prod_{k=1}^L 2\sinh (\beta\epsilon_k^{(p)}) \right] \nonumber \\
    &= \frac{1}{2} \left[ \prod_{k=1}^L e^{\beta \epsilon_k^{(p)}} (1+e^{-2\beta \epsilon_k^{(p)}}) + \prod_{k=1}^L e^{\beta \epsilon_k^{(p)}} (1-e^{-2\beta \epsilon_k^{(p)}}) \right] \nonumber \\
    &= e^{-\beta E_{vac}^{(p)}} \left( 1 + \sum_{k_1 < k_2} e^{-2\beta(\epsilon_{k_1}^{(p)} + \epsilon_{k_2}^{(p)})} + \sum_{k_1 < k_2 < k_3 < k_4} e^{-2\beta(\epsilon_{k_1}^{(p)} + \epsilon_{k_2}^{(p)} + \epsilon_{k_3}^{(p)} + \epsilon_{k_4}^{(p)})} + \dots \right),
\end{align}
where $E_{vac}^{(p)} = -\sum_{k} \epsilon_k^{(p)}$ represents the vacuum (ground state) energy of the sector.

The spectral form factor is determined by the modulus square of the partition function, which expands as
\begin{align}
    |Z|^2 &= (Z_0 + Z_1)(Z_0^* + Z_1^*) \nonumber \\
    &= |Z_0|^2 + |Z_1|^2 + 2\mathrm{Re}(Z_0 Z_1^*).
\end{align}
This expression consists of the intra-sector terms, $|Z_0|^2 + |Z_1|^2$, and the inter-sector term, $2\mathrm{Re}(Z_0 Z_1^*)$. In the low-temperature limit ($\beta_r \gg 1$), $Z_p$ is approximated by its leading excitation terms:
\begin{equation}
    Z_p \approx e^{-\beta E_{vac}^{(p)}} \left( 1 + \sum_{k_1 < k_2} e^{-2(\beta_r + i\beta_i) (\epsilon_{k_1}^{(p)} + \epsilon_{k_2}^{(p)})} \right).
\end{equation}
Evaluation of the modulus square for each sector yields
\begin{align}
    |Z_p|^2 &= Z_p Z_p^* \nonumber \\
    &\approx e^{-2\beta_r E_{vac}^{(p)}} \left( 1 + \left| \sum_{k_1 < k_2} e^{-2\beta (\epsilon_{k_1}^{(p)} + \epsilon_{k_2}^{(p)})} \right|^2 + 2 \mathrm{Re} \left[ \sum_{k_1 < k_2} e^{-2(\beta_r + i\beta_i) (\epsilon_{k_1}^{(p)} + \epsilon_{k_2}^{(p)})} \right] \right).
\end{align}
The dominant contribution to the oscillatory sum in $|Z_p|^2$ arises from terms where the single-particle energies $\epsilon_k^{(p)}$ are near their minimum, that is $\epsilon_{k}^{(p)} = \min(\epsilon_k)$. Since the spectra $\epsilon_{k}^{(p)}$ of different parity sectors coincide as $L \to \infty$, with $4\epsilon_k^{(p)}\approx \Delta_{\infty}$, the intra-sector oscillations in $S$ are primarily characterized by
\begin{equation}
    \mathrm{Re} [e^{-(\beta_r + i\beta_i)\Delta_{\infty}}] = e^{-\beta_r \Delta_{\infty}} \cos(\Delta_{\infty} \beta_i).
\end{equation}

Regarding the inter-sector contribution $2\mathrm{Re}(Z_0 Z_1^*)$, at sufficiently low temperatures, it is sufficient to retain only the ground state terms of each sector:
\begin{equation}
    Z_0 \approx e^{-\beta E_{vac}^{(0)}}, \quad Z_1 \approx e^{-\beta E_{vac}^{(1)}}.
\end{equation}
Consequently, the inter-sector term becomes
\begin{align}
    \label{eq.ReZ0Z1}
    2\mathrm{Re}(Z_0 Z_1^*) &\approx 2\mathrm{Re} \left( e^{-(\beta_r + i\beta_i) E_{vac}^{(0)}} \cdot e^{-(\beta_r - i\beta_i) E_{vac}^{(1)}} \right) \nonumber \\
    &= 2\mathrm{Re} \left( e^{-\beta_r(E_{vac}^{(0)} + E_{vac}^{(1)})} e^{-i\beta_i(E_{vac}^{(0)} - E_{vac}^{(1)})} \right).
\end{align}
As established in Ref.~\cite{SOkuyama2015}, the finite-size fine-structure gap $\Delta_L$ is defined by the energy difference between the vacuum states of the two sectors: $\Delta_L = |E_{vac}^{(0)} - E_{vac}^{(1)}|$. By setting the reference point $E_{GS} = \min (E_{vac}^{(0)}, E_{vac}^{(1)})$, Eq.~\ref{eq.ReZ0Z1} can be expressed as
\begin{equation}
    2\mathrm{Re}(Z_0 Z_1^*) \approx 2 e^{-2\beta_r E_{GS}} e^{-\beta_r \Delta_L} \cos(\Delta_L \beta_i),
\end{equation}
which generates the oscillation at frequency $\Delta_L$.

\section{Scaling invariance of giant bubbles of Fisher zeros}

\begin{figure}[t]
\includegraphics[width=0.5\textwidth]{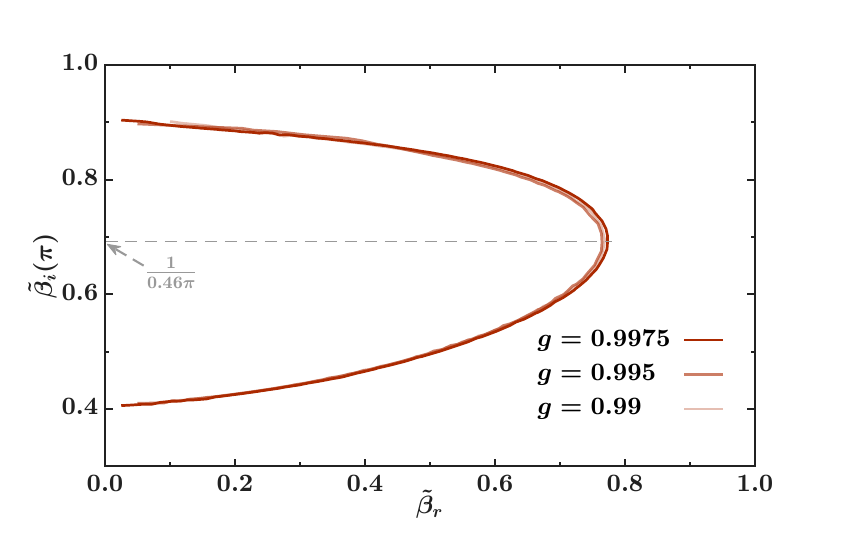}
\caption{
    Rescaled giant bubbles of Fisher zeros. Rescaled by the factor $1-g$, the complex plane coordinates are given by $\tilde{\beta}_r = (1-g) \beta_r$ and $\tilde{\beta}_i = (1-g) \beta_i$. The inverse of the $\tilde{\beta}_i$ coordinate at the rightmost zero equals 0.46.
}
\label{fig:figS1}
\end{figure}

The rightmost endpoints of the giant bubbles of Fisher zeros exhibit a linear scaling behavior, which is essentially a manifestation of the scaling invariance about $1-g$ of these giant bubbles in the complex $\beta$ plane. 
To better illustrate this point, we rescale $\beta$ as $\tilde{\beta}=(1-g)\beta$ and find that the zero-bubble contours for $g$ close to 1 nearly collapse onto each other (Fig.~\ref{fig:figS1}). Correspondingly, the linear fitting coefficient 0.46 in the main text is precisely the inverse of the imaginary part of the rightmost point in the rescaled coordinate. Within the framework of complex-$\beta$ renormalization group analysis~\cite{SMeng2025a}, the fixed zero in the thermodynamic limit can be interpreted as an unstable fixed point.

\section{Supplementary References}
%


\end{document}